\pdfoutput=1

\documentclass{sig-alternate-05-2015}
\usepackage{dsfont}
\usepackage{stmaryrd}
\usepackage{enumitem}

\newcommand{\argmax}{\arg\!\max}
\begin{document}

\setcopyright{acmcopyright}



\conferenceinfo{KDD '17}{August 13--17, 2017, Halifax, Nova Scotia, Canada}


%
\conferenceinfo{KDD '17}{August 13--17, 2017, Halifax, Nova Scotia, Canada}

\title{Real-Time Optimization Of A Web Publisher RTB Revenue}

\numberofauthors{4}
\author{
\alignauthor
        Pedro Chahuara\\
       \affaddr{XRCE}\\
       \email{pedro.chahuara@\\xrce.xerox.com}
\alignauthor
        Nicolas Grislain\\
       \affaddr{AlephD}\\
       \email{ng@alephd.com}
\alignauthor
        Gregoire Jauvion\\
       \affaddr{AlephD}\\
       \email{gj@alephd.com}
\and
\alignauthor
        Jean-Michel Renders\\
       \affaddr{XRCE}\\
       \email{jean-michel.renders@\\xrce.xerox.com}
}

\date{27 January 2017}

\maketitle
\begin{abstract}
This paper describes an engine to optimize web publishers revenue from second-price auctions, which are widely used to sell online ad spaces in a mechanism called real-time bidding. This problem is crucial for web publishers, because setting appropriate reserve prices can increase significantly their revenue.

We consider a practical setting where the only available information before an auction occurs consists of a user identifier and an ad placement identifier. Once the auction has happened, we observe censored outcomes : if the auction has been won (i.e the reserve price is smaller than the first bid), we observe the first bid and the closing price of the auction, otherwise we do not observe any bid value.

The engine predicts an optimal reserve price for each auction and is based on two key components: (i) a non-parametric regression model of auction revenue based on dynamic, time-weighted matrix factorization which implicitly builds adaptive users' and placements' profiles; (ii) a non-parametric model to estimate the revenue under censorship based on an on-line extension of the Aalen's Additive Model.

An engine very similar to the one described in this paper is applied to hundreds of web publishers across the world and yields a very significant revenue increase. The experimental results on a few of these publishers detailed in this paper show that it outperforms state-of-the-art methods and that it tackles very efficiently the censorship issue.
\end{abstract}

%
%


%
%

%
%



\section{Introduction}
Real-time bidding is a mechanism widely used by web publishers to sell their ad spaces: advertisers can bid in an auction, and the one bidding the most wins the auction and displays its ad space. The auction mechanism typically used is a second-price auction mechanism, where the winning advertiser pays the second bid.

Publishers set a reserve price, also known as floor price, under which it chooses not to sell the ad space. If all bids are under the reserve price, the ad space is not sold. Otherwise, the first bid wins the auction and the publisher revenue is the maximum between the second bid and the reserve price.

If $f$, $b_1$ and $b_2$ are respectively the reserve price and the two first bids of the auction, the revenue for the publisher is:
\begin{equation}
R(f,b_1,b_2) = \mathds{1}_{f \leq b_1} max(f,b_2)
\end{equation}

Most publishers set the reserve price levels manually on a daily or weekly basis. We propose in this paper an engine able to predict an optimal reserve price in real time for each auction, using the most fresh and relevant data.

We consider the practical setting where the only available information before the auction takes place consists of identifiers of the internet user and of the ad placement. Once the auction has happened, we observe censored outcomes : if the auction has been won (i.e the reserve price is smaller than the first bid), we observe the first bid and the closing price of the auction, otherwise we do not observe any bid value. This setting gives two different cases of censorship: (i) if the auction is lost (ie no bid above the reserve price), nothing is observed; (ii) if the reserve price is paid (closing price = reserve), then the second bid is not observed. This incomplete information setting is observed frequently in the industry.

The following constraints make this prediction problem very specific and have influenced the way the engine has been built :
\begin{itemize}
\item The bids distributions may vary significantly over time, due to the specific behavior of each advertiser, which depends for example on the data sources it uses, its budget constraint or its bidding algorithm. Thus, the engine uses time-adaptive users' and placements' profiles.
\item A web publisher has typically several thousands of different ad placements to sell and possibly several hundreds of millions of different visitors. This dimensionality limits the quantity of information stored for each placement and internet user.
\item The information available for each user may be very sparse, as a lot of them come only a few times on the publisher websites. The engine must be able to perform a relevant prediction after a few observations on a user.
\item The reserve price prediction for each auction should be taken in typically less than 10 ms.
\item Finally, the engine must tackle the censorship effects described above.
\end{itemize}

Note that we have some kind of ``closed-loop'' setting when trying to simultaneously learn the optimal reserve price strategy and, at the same time, compensating for the censored information that could result from this strategy. There is an underlying exploration/exploitation trade-off to be controlled, as a too high reserve price could prevent the model to be updated and improved (because the lack of observed output in case of censorship), even if it leads to a higher revenue, as estimated by the current model.

The engine developed in this paper performs a real-time prediction of the optimal reserve price, depending on the internet user and the ad placement (adding other features, such as the time of the day or the user operating system, is straightforward). This model is estimated on-line on possibly censored data, and the parameters are time-adaptive.

A noticeable advantage of the model used is that its use is not limited to the second-price auction mechanism : its extension to any auction mechanism where the outcome of an auction is a function of the bids is straightforward. This feature of the engine is used in practice, as some advertising platforms use non-standard auction mechanisms.

\section{Related work}
\label{related}

Revenue maximization of online ad spaces has received an increasing attention in the machine learning community during the last decade due to the fast growth of the online advertising industry \cite{Yuan2013}. Most of the work addressing this problem focuses on generalized second-price auctions with reserve price since it is the prevalent mechanism in industry. 

One of the most common assumptions when designing revenue maximization models is that information about the user and the placement is available, e.g. the device used by the user, her geographical location or the content type of the website. Accordingly, several works have presented regression models with this kind of information used as features to predict either the optimal floor or the highest bid (which can be used 
indirectly to fix an optimal floor). For instance, \cite{Cui2011} proposed an approach based on gradient boosting decision trees and mixture models in order to estimate the cumulative distribution function of the highest bid according to a set of targeting attributes. Also \cite{Wu2015} proposed a model to estimate the winning bid by means of a Tobit model in order to perform regression with censored data. This is one of the few works addressing the problem of censored data which has a major importance in RTB since lost auctions, even when not disclosing the values of the bids, provide valuable information about the interval in which the bids happened.  The main goal of these two works is predicting the first bid value, and the authors do not propose a strategy to set the reserve price. From a publisher perspective, however, estimating the optimal reserve price to maximize the revenue is the chief interest.

Another family of works tries to directly address the optimisation problem by transforming the original loss function (which has very bad properties, as it is non-convex and non-differentiable) into a surrogate function with nicer properties. By construction, these methods are parametric. As the most prominent work of this family, the one of \cite{Mohri14} tries to maximize the revenue by means of a linear predictor that is found by \textit{DC} (difference of convex functions) programming. In this case, full access to bid values is assumed. Later, \cite{Rudolph2016} described a parametric approach to determine optimal reserve prices in second-price RTB auctions by defining a smoothed revenue function to avoid non-differentiability. Their method estimates directly the optimal floor price through a Expectation-Maximization algorithm that can be used with several regressors. Nevertheless, this approach is not adaptive, does not consider censorship and can hardly be used in an on-line setting.

A main drawback of methods relying on models learned previously on training data is that they do not perform well under non-stationary environment, which is unfortunately the case in practice. In fact, the conditions in which they operate can become very different from those in which they were trained. 

The need of adaptive methods in RTB has led to the application of models based on ``Multi-armed Bandits'' (MAB) strategies \cite{CesaBianchi2015, Ikonomovska2015}. The model proposed by \cite{Ikonomovska2015} leverages a contextual MAB model which seems very appropriate when using features, however they do not deal with the uncertainty resulting from censored data.

A more simple adaptive approach to obtain an optimal reserve price has been proposed by \cite{Yuan2014}. They model the highest bid considering a log-normal distribution and after observing an auction the model is updated in a Bayesian fashion. However, the approach remains parametric and relies on the log-normal distribution assumption; moreover, it doesn't consider the censorship issue. In the same paper, the authors described another adaptive method, that basically maintains an optimal reserve price by increasing it by a small amount when it is lower than the first bid and by decreasing it by another small amount when it is larger than the first bid; this alternative method, despite its simplicity, turned out to give better performance than the one based on the log-normal assumption.

To summarize, the main differentiators of our engine rely on the following features: (i) it tackles time-varying environment and cold-start problem by incremental on-line model adaptation; (ii) it uses as sole features the identifiers of the user and of the ad placement and tackles the sparsity problems; (iii) it is able to solve the inherent censorship issues induced by the reserve price selection strategy.

\section{Description of the prediction engine}

\subsection{Problem Statement}\label{ProblemStatement}
The goal of the proposed method is to predict, before an auction happens, an optimal reserve price which maximizes the revenue in expectation. We assume a stream (sequence) of auctions, so that we can incrementally and adaptively learn an optimal strategy from previous auction outcomes.

For a given auction, the optimal \emph{a posteriori} strategy is to set a floor right below the \emph{first bid}:
$$ \argmax_f\mathds{1}_{f \leq b_1} max(f,b_2) = b_1$$
But the \emph{a priori} optimal floor is far from being the \emph{expected first bid}, because the cost of not perfectly predicting the first bid is very skewed -- setting a floor right below the first bid is not costly while setting a floor right above the first bid leads to a revenue equal to $0$. In other words,
$$ \argmax_f\mathds{E}\left(\mathds{1}_{f \leq b_1} max(f,b_2)\right) \neq \mathds{E}\left(b_1\right)$$
Choosing the expected first bid would be equivalent to minimizing a quadratic cost function which is completely different from the actual cost profile.

For this reason, we have chosen to adopt a non-parametric approach and model the whole revenue profile (in expectation) as a function of the floor instead of modelling the sole optimal floor.  By ``non-parametric'' model, we mean a model which does not rely on a pre-specified class of functions (for modelling the dependence of the revenue with respect to the reserve price), or to a predetermined distribution or family of distributions (for modelling the bid distributions). In our case, non-parametric models are built from a discretization of the input space. More precisely, we first discretize the floor space into $K$ bins (or levels) and we model the $K$ values of the revenue for the $K$ reserve prices, $(f^{(1)},...,f^{(K)})$.

To perform a prediction, the engine predicts the floor level that maximizes the expected revenue.

As we will see, updating the whole revenue profile to take into account the outcome of a new auction implies that we can compute the revenue for each floor level, which requires the knowledge of both the first and second highest bids ($b_1$ and $b_2$). However, such information is not always available: the information is ``fully-censored" (neither $b_1$ nor $b_2$ is observed) when the selected reserve price is higher than the first bid; the information is ``half-censored'' ($b_1$ is observed but not $b_2$) when the selected reserve price is lower than $b_1$ but larger than $b_2$.

The engine tackles censorship by modelling the first and second bid distributions. 
When the bid information is censored, the engine uses the bid distributions to compute the expected revenue for each floor level; which will then be fed to the revenue profile modeller as a ``proxy'' to the real one.

We first describe the revenue profile modeller component in section \ref{RevenueProfileModeller}, after having introduced some notations and definitions. Then, we will describe the bid distribution modeller component in section \ref{BidDistributionModeller}. The section \ref{OtherFeatures} briefly describes how to introduce other features than the user's and placement's identifiers. Finally, the section \ref{Complexity} assesses the computational complexity of the engine.
\subsection{Observable Inputs and Notations}

The observed features before an auction takes place are the internet user $u$ and the ad placement $p$ identifiers.

We note $\mathcal{D}$ a stream of auctions on which the model will be estimated, and $\mathcal{D}_t$ the set of auctions happening before time $t$. We denote by $\mathcal{D}_t^u$ and $\mathcal{D}_t^p$ the set of auctions happening before time $t$ involving user $u$ and placement $p$ respectively.

Once the auction happens, we observe the following information depending on the auction outcome: (i) if the auction has been won by an advertiser, we observe the reserve price $f$, the highest bid $b_1$ and the publisher revenue; (ii) if no advertiser has won the auction, we only know that the bids are upper-bounded by the reserve price $f$.

In the rest of this section, an auction will be denoted by $a$, and the corresponding time, internet user and ad placement will be denoted respectively by $t_a$, $u_a$ and $p_a$.

\subsection{Building and Updating Revenue Profile Models}
\label{RevenueProfileModeller}

In a nutshell, the method consists in predicting the revenue for any triplet <user $u$, placement $p$, reserve price level $f^{(k)}$> by latent factor decomposition.

More precisely, for $k \in \llbracket 1,K \rrbracket$, the revenue $R^{(k)}$ when the reserve price $f^{(k)}$ has been set in an auction for an internet user $u$ and an ad placement $p$ is assumed to have the following latent factor decomposition:
\begin{equation}
R^{(k)} = \beta^{(k)} + (X_{u}^{(k)})' Y_{p}^{(k)} + \epsilon^{(k)}
\end{equation}

where $\beta^{(k)}$ is a global bias for reserve price level $f^{(k)}$; $X_{u}^{(k)}$ and $Y_{p}^{(k)}$ are latent factors columns (of size $L$) associated  respectively to user $u$ and placement $p$ for reserve price level $f^{(k)}$; $\epsilon^{(k)}$ is the decomposition error term at reserve price level $f^{(k)}$ and is assumed to be gaussian with zero mean and variance $\sigma^2$. Note that the prime symbol ($'$) denotes the transpose operator;

Before an auction happens, the expected revenues $R^{(1)},...,R^{(K)}$ are predicted, and the reserve price level $f^{(k)}$ which maximizes the expected revenue is chosen for this auction. 


\subsubsection{Estimation of the Latent Factors in the Off-line Case}

Suppose that we have observed a set of auctions and their outcome over the time interval $[0,T]$, corresponding to the stream $\mathcal{D}_T$. We introduce, for $k \in \llbracket 1,K \rrbracket$, the following loss function corresponding to the estimation problem over $[0,T]$:
\begin{equation}
\label{offline1}
\begin{aligned}
L(\hat{\beta}^{(k)}, \hat{X}^{(k)}, \hat{Y}^{(k)}) =  \sum_{a \in \mathcal{D}_T} \gamma^{(T-t_a)} (R_a^{(k)} - \hat{\beta}^{(k)} - (\hat{X}_{u_a}^{(k)})' \hat{Y}_{p_a}^{(k)} )^2 \\
+ \sum_{u \in \mathds{U}} \lVert \hat{X_u}^{(k)} - X_0^{(k)} \rVert _\Omega ^2  + \sum_{p \in \mathds{P}} \lVert \hat{Y_p}^{(k)} - Y_0^{(k)} \rVert _\Gamma ^2 +  \lVert \hat{\beta}^{(k)}-\beta_0^{(k)} \rVert _\Sigma ^2
\end{aligned}
\end{equation}

where:
\begin{itemize}[leftmargin=*]
\itemsep0em 
\item the hat symbol ($\hat{.}$) denotes a parameter estimate at time $T$;
\item $\mathds{U}$ and $\mathds{P}$ are the sets of all users and placements observed in the data. We note $U$ and $P$ their sizes;
\item $\hat{X}^{(k)}$ and $\hat{Y}^{(k)}$ are the matrices of estimated latent factors for users and placements respectively (of size $U \times L$ and $P \times L$);
\item $\gamma$ is a forgetting factor ($\gamma<1$). A smaller value for the $\gamma$ hyper-parameter implies a greater time-adaptivity for the latent factors;
\item $R_a^{(k)}$ is the ``simulated'' publisher revenue when setting the $k^{th}$ reserve price level (i.e. $f^{(k)}$) in auction $a$. If the two highest bids of auction $a$, $b_1$ and $b_2$, are observed, it is simply equal to $\mathds{1}_{f^{(k)} \leq b_1} max(f^{(k)},b_2)$, otherwise it is estimated by the method described in section \ref{BidDistributionModeller};
\item $\lVert . \rVert^2_A$ is the squared Mahalanobis norm with respect to a covariance matrix $A$
\item The last three terms are regularization terms. The hyper-parameters $X_0^{(k)}$ and $\Omega$ have a direct bayesian interpretation: assuming that the user latent factors have gaussian priors of mean $X_0^{(k)}$ and covariance matrix $\Omega'$, then $\Omega$ is nothing else than the covariance matrix scaled by the inverse observation noise variance: $\Omega = \frac{\Omega'}{\sigma^2}$. $Y_0^{(k)}$, $\Gamma$, $\beta_0^{(k)}$  and $\Sigma$ have a similar interpretation. 
\end{itemize}

It is worth noting that the regularization of the problem is more general than the usual formulation where $X_0^{(k)}=Y_0^{(k)}=0$ and $\Omega=\Gamma=\lambda I$. In particular, biases per user and per placement can be estimated with this formulation. For example, if the $i^{th}$ component of the latent factor $\hat{X_u}^{(k)}$  is dedicated to capturing the bias of user $u$ at level $k$, the $i^{th}$ component of $Y_0^{(k)}$ will be set to $1$ and the $i^{th}$ diagonal value in $\Gamma$ is set at (approximately) $0$. Conversely,  if the $j^{th}$ component of the latent factor $\hat{Y_p}^{(k)}$  is dedicated to capturing the bias of placement $p$ at level $k$, the $j^{th}$ component of $X_0^{(k)}$ will be set to $1$ and the $j^{th}$ diagonal value in $\Omega$ is set at (approximately) $0$.

In the sake of notation simplicity, we drop the $(k)$ superscript here after, knowing that the following equations hold for each level.

The latent factor estimation is performed using an alternating least-squares method, which consists in estimating iteratively the factors associated to users assuming that the factors associated to placements are fixed, and reciprocally. More precisely, it amounts to iterate the following equations:
for all $u \in  \llbracket 1,U \rrbracket$, for all $p \in  \llbracket 1,P \rrbracket$, and for all $k \in  \llbracket 1,K \rrbracket$:
\begin{equation}
\label{offline1}
\begin{aligned}
\hat{X}_u & = X_0 + \left( \sum_{a \in \mathcal{D}_T^u} \gamma^{(T-t_a)} \hat{Y}_{p_a} \hat{Y}_{p_a}' + \Omega^{-1} \right) ^{-1} \\
& \left( \sum_{a \in \mathcal{D}_T^u} \gamma^{(T-t_a)} (R_a - \hat{\beta} - X_0' \hat{Y}_{p_a} ) \hat{Y}_{p_a} \right) \\
\hat{Y}_p & = Y_0 + \left( \sum_{a \in \mathcal{D}_T^p} \gamma^{(T-t_a)} \hat{X}_{u_a} \hat{X}_{u_a}' + \Gamma^{-1} \right) ^{-1} \\
& \left( \sum_{a \in \mathcal{D}_T^p} \gamma^{(T-t_a)} (R_a - \hat{\beta} - Y_0'  \hat{X}_{u_a}) \hat{X}_{u_a} \right) \\
\hat{\beta} & = \beta_0 + \left( \sum_{a \in \mathcal{D}_T} \gamma^{(T-t_a)} + \Sigma^{-1} \right) ^{-1} \\
& \left( \sum_{a \in \mathcal{D}_T} \gamma^{(T-t_a)} (R_a - \hat{X}_{{u_a}}'\hat{Y}_{p_a}) \right)
\end{aligned}
\end{equation}

\subsubsection{On-line estimation of the Latent Factors}

To estimate the latent factors on-line, we need to keep in memory the following terms for each user $u$, placement $p$ and reserve price level $k$ (the superscript $k$ is still omitted):
\begin{itemize}[leftmargin=*]
\itemsep0em 
\item $X_{u,cov}$, the time-weighted ``covariance'' matrix at time $T$: $X_{u,cov} \doteq \sum_{a \in \mathcal{D}^u_T} \gamma^{(T-t_a)} \hat{Y}_{p_a} \hat{Y}_{p_a}'$. We define symmetrically $Y_{p,cov}$;
\item $X_{u,obs}$, which represents the current estimate at time $T$ of $\sum_{a \in \mathcal{D}_T^u} \gamma^{(T-t_a)} (R_a - \hat{\beta} - X_0' \hat{Y}_{p_a} ) \hat{Y}_{p_a}$. We define symmetrically $Y_{p,obs}$;
\item $\beta_{cov}$ and $\beta_{obs}$ defined respectively by $\beta_{cov} \doteq \sum_{a \in \mathcal{D}_T} \gamma^{(T-t_a)}$ and $\beta_{obs} \doteq \sum_{a \in \mathcal{D}_T} \gamma^{(T-t_a)} (R_a - \hat{X}_u'\hat{Y}_{p_a})$
\end{itemize}

With these quantities, it is easy to derive from equation \ref{offline1} the following update equations, after observing the outcome of auction $a$: as in alternating least squares, we iterate until convergence and for each reserve price level $k$:
\begin{equation}
\label{online1}
\begin{aligned}
\hat{X}_{u_a} & = X_0 + \left( \gamma^{\Delta t_{u_a}} X_{u_a,cov} + \hat{Y}_{p_a} \hat{Y}_{p_a}' + \Omega^{-1} \right) ^{-1} \\
& \left( \gamma^{\Delta t_{u_a}} X_{u_a,obs} + (R_a - \hat{\beta} - X_0' \hat{Y}_{p_a} ) \hat{Y}_{p_a} \right) \\
\hat{Y}_{p_a} & = Y_0 + \left( \gamma^{\Delta t_{p_a}} Y_{p_a,cov} + \hat{X}_{u_a} \hat{X}_{u_a}' + \Gamma^{-1} \right) ^{-1} \\
& \left( \gamma^{\Delta t_{p_a}} Y_{p_a,obs} + (R_a - \hat{\beta} - Y_0' \hat{X}_{u_a} ) \hat{X}_{u_a} \right) \\
\hat{\beta} & = \beta_0 + \left( \gamma^{\Delta t} \beta_{cov} + 1 + \Sigma^{-1} \right) ^{-1} \\
& \left( \gamma^{\Delta t} \beta_{obs} + (R_a - \hat{X}_{u_a}' \hat{Y}_{p_a} ) \right) \\
\end{aligned}
\end{equation}

where $\Delta t$, $\Delta t_{u_a}$ and $\Delta t_{p_a}$ are respectively the time intervals since the last impression, the last impression for user $u_a$ and the last impression for placement $p_a$. In practice, one or two iterations are sufficient.

Finally, the following update formulae are applied:
\begin{equation}
\begin{aligned}
X_{u_a,cov} & \leftarrow  \gamma^{\Delta t_{u_a}} X_{u_a,cov} + \hat{Y}_{p_a} \hat{Y}_{p_a}' \\
X_{u_a,obs} & \leftarrow  \gamma^{\Delta t_{u_a}} X_{u_a,obs} + (R_a^{(k)} - \hat{\beta} - X_0' \hat{Y}_{p_a}) \hat{Y}_{p_a} \\
Y_{p_a,cov} & \leftarrow  \gamma^{\Delta t_{p_a}} \times Y_{p_a,cov} + \hat{X}_{u_a} \hat{X}_{u_a}' \\
Y_{p_a,obs} & \leftarrow  \gamma^{\Delta t_{p_a}} \times Y_{p_a,obs} + (R_a^{(k)} - \hat{\beta} - Y_0' \hat{X}_{u_a}) \hat{X}_{u_a} \\
\beta_{cov} & \leftarrow \gamma^{\Delta t} \beta_{cov} + 1 \\
\beta_{obs} & \leftarrow  \gamma^{\Delta t} \beta_{obs} + (R_a-\hat{X}_{u_a}' \hat{Y}_{p_a} ) \\
\end{aligned}
\end{equation}

At each iteration and for each reserve price level, there are $2$ inversions of $L \times L$ matrices. These inversions do not lead to practical problems because $L$ is chosen relatively low in practice. 

Note that using a loss function based on the average least-squares error $\frac{1}{\sum_{a \in \mathcal{D}_T} \gamma^{(T-t_a)}} \sum_{a \in \mathcal{D}_T} \gamma^{(T-t_a)} (R_a^{(k)} - \hat{\beta}^{(k)} - (\hat{X}_{u_a}^{(k)})' \hat{Y}_{p_a}^{(k)} )^2$ would lead to much simpler update formulae with no matrix inversion (thanks to the matrix inversion lemma). However, the loss function introduced here is much more adapted to the on-line setting. Indeed, it enables to give a bayesian interpretation of the regularization: a gaussian prior is assigned to each user/placement latent factors. The more observations are available for a user or for a placement, the less the prior impacts the latent factor estimation.

\subsection{Dealing with Censored Observations}
\label{BidDistributionModeller}

When $b_1$ and $b_2$ are observed for an auction $a$, it is easy to compute or, in other words, to simulate the revenue for any floor level: $R_a^{(k)}=\mathds{1}_{f^{(k)} \leq b_1} max(f^{(k)},b_2)$. Obviously, this could not be done when the bid values are censored. This section describes how we still can estimate $R_a^{(k)}$ with censored observations. Remember that we have two kinds of censorships in the data: half and full censorship (section \ref{ProblemStatement} )

\subsubsection{A Brief Recap of the Aalen's Additive Regression Model}

Let's first recall the Aalen's regression method \cite{Aalen1989} for left-censored data, that allows us to estimate the cumulative hazard rate of a variable -- and, consequently, its cumulative distribution function (CDF) --  by means of a set of features (covariates). Note that, for left-censored data, the term ```hazard rate'' is an abuse of language, as it is defined here as the ratio of the probability density function, $pdf$, over the cumulative distribution, $cdf$, while its standard definition  is the ratio $pdf/(1-cdf)$  (the standard case considers  right-censored data). Given a discretized variable $V$ with  values in the set $(v^{(1)},...,v^{(K)})$ , $n$ observations of this variable ($v_i$ with $i=1, \ldots, n$), and a vector $C$ indicating which of these observations are (left) censored ($C_i$ is 1 if observation $i$ is non-censored and 0 when it is censored),  the hazard rate of $V$ at a level $v^{(k)}$ can be modeled as a linear combination of $(p+1)$ features represented by the vector $x=(1,x_1,...,x_p)$ (the first feature aims at capturing the bias which is commonly called the ``basis hazard rate'' in this framework):

\begin{equation}
\lambda(v^{(k)}|x)=\beta^{(k)}_0 +\beta_1^{(k)} x_1+...+\beta_p^{(k)}x_p
\end{equation}

 In order to estimate the ($(p+1) \times K$) coefficients $\beta^{(k)}_j$, we basically solve $K$ linear regression problems as follows: for $k = 1, \ldots, K$, we first select the subset $S_k$ of observations with $v_i<=v^{(k)}$ (either censored or not); we then build the corresponding feature matrix $X_k$ , of size $|S_k| \times (p+1)$, which is composed of the feature vectors $x$ of the $|S_k|$ observations and the target vector $Y_k$ of size $|S_k| \times 1$ that contains 1 for the non-censored observations with $v_i=v^{(k)}$ and 0 otherwise. The coefficients are therefore estimated using a standard regularised least-squares regression method:

\begin{equation}
\beta^{(k)}=(X_k' X_k + \lambda I)^{-1}X_k' Y_k
\end{equation}

The coefficients for the cumulative hazard rate can be estimated as:

\begin{equation}
B^{(k)}= \sum_{j \geq k}\beta^{(j)}
\end{equation}

Then, given an observation with feature vector $x$, its cumulative hazard rate at level $v^{(k)}$ is given by:

\begin{equation}
\Lambda(v^{(k)}|x)= x.B^{(k)} = B_0^{(k)}+ B_1^{(k)} x_1 + \ldots + B_p^{(k)} x_p
\end{equation}

The CDF of the variable $V$ with censored observations can finally be obtained as:

\begin{equation}
\Phi(v^{(k)}|x)= \exp(-\Lambda(v^{(k)}|x))
\end{equation}

\subsubsection{Using Aalen's regression model to estimate first and second bids' distribution}

Let's now consider how Aalen's regression model could be used for solving the issue of left-censored observations in the reserve price optimization problem. In a nutshell, two Aalen's regression models will continuously and adaptively provide the engine with an estimate of the distribution of the first and the second bid distributions independently. At any moment, the estimation of both bid probability distributions can be used in order to estimate the expected revenue for different values of the reserve price. Once again, the models are not parametric, in the sense that they do not assume any prior distribution, unlike several state-of-the-art approaches that assume a log-normal distribution for bids. So, we work on a discretized bid space, using $K'$ bins (or levels): $(b^{(1)}, b^{(2)}, \ldots, b^{(K')})$. Even if this is not required, we will assume for simplicity that $K=K'$ and that the set of discretized values for $b_1$ and $b_2$ is the same as for the reserve price ($f$). Basically, the Aalen's method provides an estimate of the CDF at each of these values. Remarkably, the way the engine maintains probability distributions over the bids is very similar to the way the engine is maintaining a revenue estimation for the different levels of floor: the updates are done for the different discretized levels using a decomposition into latent factors related either to the user or to the placement (but not to the particular $<user,placement>$ pair, avoiding some severe sparsity issues). Moreover, the update equations have actually the same form. 

When applying the Aalen's additive model to estimate the first bid distribution, we will assume a latent factor model of the following form: for $k \in \llbracket 1,K \rrbracket$, and for an auction for an internet user $u$ and an ad placement $p$
\begin{equation}
\lambda_1^{(k)} =  (M_{u}^{(k)})' N_{p}^{(k)} + \eta^{(k)}
\end{equation}
where $\lambda_1^{(k)}$ is the hazard rate of the first bid distribution at level $k$ (bins $b^{(k)}$); $M_{u}^{(k)}$ and $N_{p}^{(k)}$ are latent factors columns (of size $L$) associated  respectively to user $u$ and placement $p$ for first bid level $b^{(k)}$; $\eta^{(k)}$ is the decomposition error term at first bid level $b^{(k)}$ and is assumed to be gaussian with zero mean and variance $\sigma_1^2$.

All the update equations that will be described hereafter for the first bid distribution are identical for the second bid, except for some details which we will mention explicitly.

As we did for the revenue latent factor estimation, we introduce, for $k \in \llbracket 1,K \rrbracket$, the following loss function for the estimation problem over $[0,T]$:
\begin{equation}
\begin{split}
L(\hat{M}^{(k)}, \hat{N}^{(k)}) = \sum_{a \in \mathcal{D}_T^{(k)}} \gamma_1^{(T-t_a)} (C_a^{(k)} - (\hat{M}_{u_a}^{(k)})' \hat{N}_{p_a}^{(k)} )^2 \\ 
+ \sum_{u \in \mathds{U}} \lVert \hat{M_u}^{(k)} - M_0^{(k)} \rVert _{\Omega_1} ^2 + \sum_{p \in \mathds{P}} \lVert \hat{N_p}^{(k)} - N_0^{(k)} \rVert _{\Gamma_1}^2 \\
+ \lVert \eta^{(k)}-\eta_0^{(k)} \rVert ^2
\end{split}
\end{equation}
\\
where:
\begin{itemize}[leftmargin=*]
\itemsep0em 
\item $\mathcal{D}^{(k)}_T$ is the set of auctions up to time $T$ whose first bid or its left-censored value is smaller or equal to level $b^{(k)}$; in other words, this is the set of historical auctions for which $\max(b_1,f_a) \leq b^{(k)}$ ($f_a$ is the floor for auction $a$);
\item $C_a^{(k)}=1$ if the first bid is uncensored AND if the bid belongs to the bin $b^{(k)}$; $C_a^{(k)}=0$ otherwise;
\item $\hat{M}^{(k)}$ and $\hat{N}^{(k)}$ are the matrices of estimated first bid latent factors for users and placements respectively (of size $U \times L$ and $P \times L$);
\item $\gamma_1$ is a forgetting factor ($\gamma_1<1$);
\item The last two terms are regularization terms and have a direct bayesian interpretation (section \ref{RevenueProfileModeller}).
\end{itemize}

Note that, if we consider the second bid distribution, these definitions should be adapted in the following way:
\begin{itemize}[leftmargin=*]
\itemsep0em 
\item $\mathcal{D}^{(k)}_T$ is the set of auctions up to time $T$ whose second bid or its left-censored value is smaller or equal to level $b^{(k)}$; in other words, this is the set of historical auctions for which $\max(b_2,f_a) \leq b^{(k)}$ ($f_a$ is the floor for auction $a$);
\item $C_a^{(k)}=1$ if the second bid is uncensored and if this bid belongs to the bin $b^{(k)}$; $C_a^{(k)}=0$ otherwise;
\end{itemize}

As before, the latent factor  estimation is performed using an alternating least-squares method. We directly give the update equations for the on-line setting, as they constitute the core of the proposed methods.

So, in order to estimate the latent factors on-line, we need to keep in memory the following terms for each user $u$, placement $p$ and bid level $k$ :
\begin{itemize}[leftmargin=*]
\itemsep0em 
\item $M^{(k)}_{u,cov}$, the time-weighted ``covariance'' matrix at time $T$: $M^{(k)}_{u,cov} \doteq \sum_{a \in \mathcal{D}^{u,(k)}_T} \gamma^{(T-t_a)}  \hat{N}^{(k)}_{p_a} \hat{N}^{(k)'}_{p_a}$. We define symmetrically $N^{(k)}_{p,cov}$. Note the difference with respect to equation \ref{online1}: the sum is over the set $\mathcal{D}^{u,(k)}_T$, meaning the set of all historical auctions where user $u$ appeared and for which $\max(b_1,f_a) \leq b^{(k)}$;
\item $M^{(k)}_{u,obs}$, which represents the current estimate at time $T$ of $\sum_{a \in \mathcal{D}_T^{u,(k)}} \gamma^{(T-t_a)} (C_a^{(k)}  - M_0^{(k)'} \hat{N}^{(k)}_{p_a} ) \hat{N}^{(k)}_{p_a}$. We define symmetrically $N^{(k)}_{p,obs}$.
\end{itemize}

With these quantities, the following update equations, after observing the outcome of auction $a$, are iterated until convergence and for each bid level $k$:
\begin{equation}
\label{online2}
\begin{aligned}
\hat{M}^{(k)}_{u_a} & = M_0^{(k)} + \left( \gamma_1^{\Delta t^{(k)}_{u_a}} M^{(k)}_{u_a,cov} + \hat{N}^{(k)}_{p_a} \hat{N}^{(k)'}_{p_a} + \Omega_1^{-1} \right) ^{-1} \\
& \left( \gamma_1^{\Delta t_{u_a}} M^{(k)}_{u_a,obs} + (C_a^{(k)} - M^{(k)'}_0 \hat{N}^{(k)}_{p_a} ) \hat{N}^{(k)}_{p_a} \right) \\
\hat{N}^{(k)}_{p_a} & = N_0^{(k)} + \left( \gamma_1^{\Delta t^{(k)}_{p_a}} N^{(k)}_{p_a,cov} + \hat{M}^{(k)}_{u_a} \hat{M}^{(k)'}_{u_a} + \Gamma_1^{-1} \right) ^{-1} \\
& \left( \gamma_1^{\Delta t_{p_a}} N^{(k)}_{p_a,obs} + (C_a^{(k)}  - N^{(k)'}_0 \hat{M}^{(k)}_{u_a} ) \hat{M}^{(k)}_{u_a} \right) \\
\end{aligned}
\end{equation}

where  $\Delta t^{(k)}_{u_a}$ and $\Delta t^{(k)}_{p_a}$ are respectively the time intervals since the last impression for user $u_a$ and the last impression for placement $p_a$, restricted to past auctions for which $\max(b_1,f_a) \leq b^{(k)}$ . In practice, one or two iterations are sufficient.

Finally, the following update formulae are applied:
\begin{equation}
\begin{aligned}
M^{(k)}_{u_a,cov} & \leftarrow  \gamma_1^{\Delta t^{(k)}_{u_a}} M^{(k)}_{u_a,cov} + \hat{N}^{(k)}_{p_a} \hat{N}^{(k)'}_{p_a} \\
M^{(k)}_{u_a,obs} & \leftarrow  \gamma_1^{\Delta t^{(k)}_{u_a}} M^{(k)}_{u_a,obs} + (C_a^{(k)} - M_0^{(k)'} \hat{N}^{(k)}_{p_a}) \hat{N}^{(k)}_{p_a} \\
N^{(k)}_{p_a,cov} & \leftarrow  \gamma_1^{\Delta t^{(k)}_{p_a}} \times N^{(k)}_{p_a,cov} + \hat{M}^{(k)}_{u_a} \hat{M}^{(k)'}_{u_a} \\
N^{(k)}_{p_a,obs} & \leftarrow  \gamma_1^{\Delta t^{(k)}_{p_a}} \times N^{(k)}_{p_a,obs} + (C_a^{(k)} - N_0^{(k)'} \hat{M}^{(k)}_{u_a}) \hat{M}^{(k)}_{u_a} \\
\end{aligned}
\end{equation}

\subsubsection{Revenue Profile Estimation from the Estimated Bid Distributions}

We can derive the CDF of $b_1$ and $b_2$, from their hazard rates: for instance, the CDF of $b1$, denoted by $\Phi_1(b^{(k)})$ can be computed as: $\Phi_1(b^{(k)}|u,p) = \exp(-\sum_{j \geq k} \lambda_1^{(j)}(u,p))$. The CDF of $b_1$ and $b_2$ can then be used to estimate the expected revenue at each discrete value of the floor price and this estimate is then used to feed the Revenue Profile Modeller.

Let's first consider the full censorship case ($b_2 \leq b_1 \leq f_a$ where $f_a$ is the floor selected for the current auction $a$). Since the revenue is given by $R(f,b_1,b_2)= b_2 \mathds{1}_{b_2 > f} +f \mathds{1}_{b_2 \leq f \leq b_1} $, the expected revenue in the continuous case can be estimated as:
\begin{equation}
\begin{aligned}
& \mathbb{E}(R(f,b_1,b_2 | b_2 \leq b_1 \leq f_a)) = \\
& \int_{f}^{f_a} P(b_2>t | b_2 \leq b_1 \leq f_a)dt + f P(f \leq b_1 | b_2 \leq b_1 \leq f_a) \\
& \simeq \int_{f}^{f_a} P(b_2>t | b_2 \leq f_a)dt + f P(f \leq b_1 | b_2 \leq b_1 \leq f_a)
\end{aligned}
\end{equation}
Details of the derivation can be found in equation (2) of  \cite{Mohri14}. The second equation emphasizes the approximation that allows us to model independently the first and second distributions, instead of modelling the joint distributions. Experimentally, this approximation could be shown to have virtually no impact on the computation of the expected revenue, at least with the auction datasets we used.

This second equation can be converted for the discrete case into the following equation:
\begin{equation}
\begin{aligned}
& \forall f^{(k)} < f_a , \quad \mathbb{E}_{b_1,b_2}(R(f^{(k)},b_1,b_2 | b_2 \leq b_1 \leq f_a))= \\
& \sum_{k'=k}^{f_a} f^{(k')} \tilde{\phi_2} (b^{(k')}) - f^{(k)}.(1-\tilde{\Phi_2}(f^{(k)})) +f^{(k)}.(1-\tilde{\Phi_1}(f^{(k)})) \\
\end{aligned}
\end{equation}
where  $\tilde{\Phi_1}$ and  $\tilde{\Phi_2}$ are the re-normalised c.d.f.'s of  the first and second bid respectively, conditioned by the fact that they should be smaller than $f_a$: $\tilde{\Phi_1}(f)=\frac{\Phi_1(f)}{\Phi_1(f_a)}$, $\tilde{\Phi_2}(f)=\frac{\Phi_2(f)}{\Phi_2(f_a)}$ with
$\Phi_1$ and $\Phi_2$  the (non-conditional) c.d.f.'s of the first and second bid respectively, while $\tilde{\phi_2}$ is the (discrete) renormalised p.d.f of the second bid corresponding to the $\tilde{\Phi_2}$ CDF. For levels $f^{(k)} \geq f_a$, the revenue is equal to 0.

For the half-censored case ($b_1$ observed and $f_a \leq b_1$), this formula becomes:
\begin{equation}
\begin{aligned}
& \mathbb{E}_{b_2}(R(f^{(k)},b_1,b_2 | b_2 \leq f_a))= \\
& \sum_{k'=k}^{f_a} f^{(k')} \tilde{\phi_2} (b^{(k')}) - f^{(k)}.(1-\tilde{\Phi_2}(f^{(k)})) +f^{(k)} \quad \forall f^{(k)} < f_a
\end{aligned}
\end{equation}

For levels $f^{(k)} \geq f_a$, the revenue is not ``censored'' and is equal to $f^{(k)}$ if $f^{(k)} \leq b_1$ and 0 otherwise.

\subsection{Introduction of additional features}
\label{OtherFeatures}

We detail briefly how to introduce contextual features other than the user's and placement's identifier into the model , e.g. the time of the day or the user's device. Let's $\theta$ be the corresponding feature vectors.  For the sake of simplicity, we assume that the dependency between the feature vector and the revenue is linear and that it simply combines additively with the biases and latent factors:
\begin{equation}
R^{(k)} = \beta^{(k)} + (X_{u}^{(k)})' Y_{p}^{(k)} + \theta' Z^{(k)} + \epsilon^{(k)}
\end{equation}

In this setting, the formulae to estimate $\beta^{(k)}$ and the latent factors $X_{u}^{(k)}$ and $Y_{p}^{(k)}$ are kept unchanged, except that $R^{(k)}$ is replaced by $(R^{(k)} - \theta' Z^{(k)})$.
It is straightforward to derive the formulae to estimate adaptively the parameters $Z^{(k)}$: it is similar to the update equations of $\beta^{(k)}$, except that it uses $\theta$ as regressor, instead of the \textbf{1} constant.

\subsection{Computational complexity of the engine}
\label{Complexity}

Let $K$ be the number of reserve price levels, $L$ be the size of the latent factors space and $I$ be the number of iterations performed when updating the latent factors.

The most computationally expensive operations are the updates of the latent factors used in the revenue profile and bid distribution modellers. These updates require a $L \times L$ matrix inversion for each iteration and for each reserve price level, i.e $K \times I$ matrix inversions.

In practice, an appropriate value for $K$ is around $100$, and $5$ for $I$. $L$ is also low (see section \ref{Results}), which makes the engine suitable in real applications.

\section{Results}\label{Results}

In this section we describe results obtained from comparing our method with baseline approaches that are appropriate for the task and also with state-of-the-art methods that have been proposed recently in the literature.

The evaluation is performed on a real advertising publisher dataset containing more than $4.4$ millions of auctions data collected over one week. It contains 367K unique users and 2K unique placements, as well as the first and second bid values (non-censored) of each auction which provides the ground truth to evaluate our method with respect to the full information (non-censored) setting.  

The dataset is divided into two subsets: the training set (containing the first 3 days of observations, approximately 3/7 of the total number of observations) and the test set. The results given below are estimated on the test set.

For all the evaluations, the metric to measure peformance is the resulting revenue. It is a reasonable measure since the primary goal is to maximize the revenue, conversely to other works \cite{Wu2015} which try to predict the highest bid in which case the estimation error is a more appropriate metric.  

Initially, the revenue achieved when applying the reserve price predicted by the model is compared to the  following two reserve price setting baselines:
\begin{itemize}[leftmargin=*]
\itemsep-0.2em 
\item $NO\_RES$: reserve price equal to $0$;
\item $PL\_RES$: setting an optimal reserve price per placement determined on the training set (using uncensored bid values);
\item $PL\_RES\_ONLINE$: setting an optimal reserve price per placement which is estimated online as the reserve price maximizing an exponentially-weighted moving average of the revenue (using uncensored bid values)
\end{itemize}

We consider 3 settings:

\textbf{S1:} The ideal one (or easy one), where the training is uncensored (but the test set is censored); at the end of the training phase, as usual, the identified values of the latent factors are used as initial values when starting the test phase;

\textbf{S2:} The hard one, where both the training and test sets are censored; note that the censorship of the training set is fixed to the historical prices and could not be modified, while the censorship level in the test set could be controlled by the reserve price optimisation strategy (we use the knowledge of both $b_1$  and $b_2$ to simulate the revenue of the proposed strategies); moreover, we assume that the training set could be used only as a ``development set'' to tune the hyper-parameters, but not to initialise the values of the latent factors (as if a reset operation has been applied just before the test set); the goal of this constraint is to emphasize the ``cold-start'' and adaptive capabilities of our algorithm;

\textbf{S3:} An intermediate one, where only the test set is censored; but here also, we assume that the training set could be used only as a ``development set'' to tune the hyper-parameters, but not to initialise the values of the latent factors.

Besides the two baselines, we consider 4 variants of our method:

\textbf{M1:} The complete one, using both the revenue profile modeller and the bid distribution modeller when the bid information is censored

\textbf{M2:} A variant where only the revenue profile modeller is used and, to handle the censorship issue, we arbitrarily fix $b_1$ and $b_2$ to 0 in case of full censorship, and $b_2$ equal to $f_a$ in case of half-censorship; intuitively, this amounts to never favouring ``higher'' levels of floor in case of censorship so that we can promote exploration of low levels of floor (remember that, when we know the revenue for a floor, we automatically know the revenue for all floors that are larger than this floor). So, this is an indirect way of controlling the exploration/exploitation trade-off;

\textbf{M3:} A variant where only the revenue profile modeller is used and, to handle the censorship issue, we simply skip it, meaning that we do not update the revenue profile for the part which is unknown due to censorship (but we update it for all levels that are larger or equal to the current reserve price);

\textbf{M4:} the same variant as \textbf{M3}, but instead of selecting the reserve price whose predicted expected revenue is the largest, we consider the Lin-UCB selection strategy.

Only (\textbf{M1}) uses the bid distribution modeller component.

In the update equations of the adaptive methods, the parameters (biases and latent factors) were initialised to small random values (gaussian with 0.1 variance).

The performances of the different methods, namely the average revenues on the Test Set (in arbitrary monetary units) are given in the following table:

\bigskip

\begin{tabular}{p{3cm} c c c}
\cline{2-4}
 & S1 & S2 & S3 \\ \hline
NO\_RES & 2.5978 & N/A & N/A \\
PL\_RES & 3.6222 & N/A & N/A \\ 
PL\_RES\_ONLINE & 3.7154 & N/A & N/A \\ \hline
M1 & 4.0663 & 3.9955 & 3.9957 \\
M2 & 3.9012 & 3.7948 & 3.7936 \\
M3 & 3.8954 & 3.7369 & 3.7468 \\
M4 & 3.9306 & 3.821 & 3.8209 \\ \hline
Oracle (knowing $b_1$) & & 8.6552 & \\ \hline
\end{tabular}

\bigskip

Some remarks about the results and the choice of the hyper-parameters:
\begin{itemize}[leftmargin=*]
\itemsep-0.2em 
\item Starting the model from scratch at the beginning of the test set finally has a small impact on the performance; this is not surprising, knowing that there are a lot of ``flash'' users (i.e. new users who never appeared before and will disappear a few minutes later) for which solving the cold-start problem is crucial;
\item A censored training set is not detrimental to performance: results are nearly the same than with a uncensored training set;
\item In the case of both training and test sets uncensored (so that no censorship-dealing strategy should be used), the revenue profile modeller gives a performance of 4.163 (4.1294 if we apply a ``reset'' operation before the test set, to emphasize cold-start performance). The method \textbf{M1} is relatively close to this level of performance and it shows that the bid distribution modeller performs well; 
\item The hyper-parameters are determined using grid-search (6 discrete values per hyper-parameter on a logarithmic scale, from $10^{-6}$ to $10^{-1}$ for the forgetting factors and from $10^{0}$ to $10^{5}$ for the constant diagonal covariance priors), focusing on the ones that give the best average revenue on the training set. Optimal results are reached when the observations are forgotten after a few minutes for users and a few hours for placements;
\item The dimension of the latent space can be kept low (in our case 2): once the latent factors corresponding to user and placement biases are added in the model, adding new latent factors does not improve very significantly the results to the price of a high complexity. This comes probably from the relatively low number of observations per user, which improves the risk of over-fitting if the dimension is too large. The optimal value for the latent space dimension is probably dependent on the dataset
\end{itemize}

Finally, we have compared our method with three state-of-the-art approaches: the one based on a Bayesian smoothing of the revenue function (the parametric approach based on an \textit{EM}-like algorithm as described in \cite{Rudolph2016}), the one based on an assumed log-normal distribution of the first bid \cite{Yuan2014} and the one based on the simple adaptation mechanism (``increase the floor when it is below the first bid; decrease it  when it is larger''), as described in \cite{Yuan2014} but extended to maintain one optimal floor value per placement. Note that the former method is not adaptive (the parameters are fixed after a training phase), while the last two methods are. Both methods assume that the bid information is uncensored. Therefore, in order to keep the comparison fair, we compared these benchmark methods with ours in the full information setting (training and test sets uncensored). The average revenues on the test set are summarized in the following table:

\bigskip

\begin{tabular}{p{6cm} c}
\cline{2-2}
 & S1 \\ \hline
Bayesian Smoothing (non-adaptive) & 2.893  \\ \hline
Log-Normal Bid Distribution (adaptive) & 2.932  \\ \hline
Simple Bi-directional Inc/Dec (adaptive) & 3.627 \\ \hline
M1 & 4.163 \\ \hline
\end{tabular}

\bigskip

The relatively poor performance of our benchmark methods most probably comes from the fact that their underlying assumptions are violated in practice. Indeed, as far as method \cite{Yuan2014} is concerned, assuming a log-normal distribution for the first bid and taking the mean minus a small constant for setting the reserve price is a too simplistic strategy, due to the high skewness of the revenue function as we explained before. For the Bayesian smoothing method of \cite{Rudolph2016}, considering that the optimal reserve price could be expressed as a linear function of the available features -- which means here that it could be expressed as the sum of a user weight and a placement weight -- seems also to be too simplistic. Note that this method uses only $P+U$ parameters (when only information available is the user identifier and the placement identifier), while our method uses $K.(P+U)$ parameters; moreover, this method does not take into account the time-varying aspects of the problem and this also explains its low performance. The simple adaptive ``increase/decrease'' method turned out to be only slightly superior to the non-adaptive $PL\_RES$ strategy, implying that the adaptation mechanism did not really succeed in capturing the time-varying properties of the first bid distribution. 

We have also measured the practical efficiency of the proposed method in terms of CPU response time using a standard computer ($\mu P$ 3.50GHz, RAM 8GB). Updating the full model(M1 above) and estimating the optimal floor price in the worst case (under full censorship) requires only 0.38 ms, and even less without censorship (0.21 ms). This is far below the limit of 10ms at which the optimal floor must be decided.
\section{Conclusions and Future Directions}

Deploying a revenue maximization engine through the optimal setting of the reserve price for real-time-bidding auctions remains a challenging industrial problem: the ``very short latency'' constraint with its implication on the algorithm complexity, the sparsity of user and placement information and, last but not least, the highly time-varying environment all raise strong issues to be solved. In this paper, we adopt a non-parametric approach to adaptively predict the revenue profile for each floor level by an on-line matrix factorisation approach. To compensate for the lack of full bid information due to the intrinsic censorship of the auction mechanism, we propose to use a similar approach for predicting the first and second bid distributions through the Aalen's additive regression model. We validated this approach by deploying the engine on a real dataset, including millions of auctions over a time period of one week: results show that the proposed approach outperforms state-of-the-art methods and, in particular, that the average revenue nearly reaches the level of the non-censored setting.

The model can be extended to deal with more complex non-linear interactions than the one assumed by a simple matrix factorization, especially when some extra user or placement features are available, keeping in mind that the real-time constraints could give some restriction on the class of models in practice.

Finally, a more global game-theoretic approach should be adopted, by considering the case where the bidders' strategies could themselves evolve over time when they observe the reserve price setting strategy: this introduces some closed-loops that makes the problem much harder to solve.

\bigskip

\textbf{Aknowledgment}: This work was partially funded by the French Government under the grant   $<$ANR-13-CORD-0020$>$ (ALICIA Project).

%
\bibliographystyle{abbrv}
\bibliography{xrce_alephd_paper}  
%
%
\end{document}